%% file: main.tex
\documentclass[a4paper,fleqn]{cas-dc}

\usepackage[numbers,sort]{natbib}
\usepackage{amsmath,amssymb,amsfonts}
\usepackage{algorithm}
\usepackage[noend]{algpseudocode}
\usepackage{graphicx}
\usepackage{subcaption}
\usepackage{xcolor}
\usepackage{todonotes}
\def\tsc#1{\csdef{#1}{\textsc{\lowercase{#1}}\xspace}}
\tsc{WGM}
\tsc{QE}
\tsc{EP}
\tsc{PMS}
\tsc{BEC}
\tsc{DE}

\begin{document}
\let\WriteBookmarks\relax
\def\floatpagepagefraction{1}
\def\textpagefraction{.001}
\shorttitle{A Case Study on Parallel HDF5 File Concatenation for HEP Data Analysis}
\shortauthors{S Lee et~al.}

\title [mode = title]{A Case Study on Parallel HDF5 Dataset Concatenation for High Energy Physics Data Analysis}

\author [1]{Sunwoo Lee}
\cormark[1]
\author [1]{Kai-yuan Hou}
\author [1]{Kewei Wang}
\author [2]{Saba Sehrish}
\author [2]{Marc Paterno}
\author [2]{James Kowalkowski}
\author [3]{Quincey Koziol}
\author [4]{Robert B. Ross}
\author [1]{Ankit Agrawal}
\author [1]{Alok Choudhary}
\author [1]{Wei-keng Liao}
\address[1]{Northwestern University, 2145 Sheridan Rd, Evanston, IL 60208, USA}
\address[2]{Fermi National Accelerator Laboratory, Kirk Rd, Batavia, IL 60510, USA}
\address[3]{Lawrence Berkeley National Laboratory, 1 Cyclotron Rd. Berkeley, CA 94720, USA}
\address[4]{Argonne National Laboratory, 9700 S Cass Ave. Lemont, IL 60439, USA}

\cortext[cor1]{Corresponding author}

\begin{abstract}
In High Energy Physics (HEP), experimentalists generate large volumes of data that, when analyzed, helps us better understand the fundamental particles and their interactions.
This data is often captured in many files of small size, creating a data management challenge for scientists.
In order to better facilitate data management, transfer, and analysis on large scale platforms, it is advantageous to aggregate data further into a smaller number of larger files.
However, this translation process can consume significant time and resources, and if performed incorrectly the resulting aggregated files can be inefficient for highly parallel access during analysis on large scale platforms.
In this paper, we present our case study on parallel I/O strategies and HDF5 features for reducing data aggregation time, making effective use of compression, and ensuring efficient access to the resulting data during analysis at scale.
We focus on NOvA detector data in this case study, a large-scale HEP experiment generating many terabytes of data.
The lessons learned from our case study inform the handling of similar datasets, thus expanding community knowledge related to this common data management task.

\end{abstract}




\maketitle

\input{1_intro}
\input{2_nova}
\input{3_workflow}

\input{4_exp_settings}
\input{5_meta}
\input{6_raw}

\input{7_eval}
\input{8_summary}
\input{9_related}
\input{10_conclusion}
\section* {Acknowledgment}
This material is based upon work supported by the U.S. Department of Energy, Office of Science, Office of Advanced Scientific Computing Research, Scientific Discovery through Advanced Computing (SciDAC) program, under the ``RAPIDS Institute'' and grant ``HEP Data Analytics on HPC'', No. 1013935. 
This work is also supported in part by the DOE awards DE-SC0014330 and DE-SC0019358.
This research was also supported by the Exascale Computing Project (17-SC-20-SC), a collaborative effort of the
U.S. Department of Energy Office of Science and the National Nuclear Security Administration through the ExaHDF5
project under Contract No. DE-AC02-05CH11231.
This research used resources of the National Energy Research Scientific Computing Center, 
a DOE Office of Science User Facility supported by the Office of Science of the U.S. Department
of Energy under Contract No. DE-AC02-05CH11231.
This research used resources of the Oak Ridge Leadership Computing Facility at the Oak Ridge National Laboratory, which is supported by the Office of Science of the U.S. Department of Energy under Contract No. DE-AC05-00OR22725.


\bibliographystyle{cas-model2-names}

\bibliography{cas-refs}

\end{document}

%% file: 1_intro.tex
\section{Introduction} \label {sec:intro}

In high energy physics (HEP), the quantities of experimental data generated from large-scale instruments are often immense, imposing demands on scalable software solutions for I/O and data analysis \cite{python,spark,eventselection}.
Traditional HEP workflows have been designed for a grid-oriented environment.
In this environment, parallelism in data reduction programs is achieved by having many independent processes handling separate files, thus producing large numbers of output files to be used for the final statistical analysis of the reduced data.
In order to better facilitate data management, transfer, and analysis on large-scale platforms, it is advantageous to aggregate the data into a smaller number of larger files.
However, the data aggregation process can consume significant time and resources.
Considering the fast growth in size of HEP experimental data, the bottleneck introduced by this translation process is becoming increasingly expensive.

In this paper we present a case study of a data analysis component of the NuMI Off-axis $\nu_e$ Appearance (NOvA) experiment that is designed to study neutrino oscillations using the particle collision event data recorded by two accelerator-based detectors \cite{nova}.
Because of their grid-oriented processing environment,
NOvA produces a large number of Hierarchical Data Format (HDF5) files.
Each HDF5 file has many groups, each of which contains many datasets.
All NOvA files have the same groups and datasets and contain the measured data from separate time periods.
We consider the data aggregation process such that, given hundreds to thousands of HDF5 files, each dataset is concatenated across all the input files and written into a single shared file.
This concatenation is a critical step enabling the NOvA data analysis component to search through the entire detector readouts in parallel and identify an often-small fraction that contains neutrino interactions of interest \cite{eventselection, spark}.

HDF5 \cite{hdf5}, a popular I/O library in scientific communities since the late 1990s \cite{python,ultra,bio,genome,scientific,scientific2,earth}, enables users to store data in a portable, self-describing file format and has started supporting parallel I/O for data compression with version 1.10.3.
In addition, many open-source and commercial software packages for data visualization and analysis read and write HDF5 files\footnote{https://portal.hdfgroup.org/display/support/Software+Using+HDF5}, making HDF5 an ideal I/O method for analysis within the NOvA experiment.

NOvA data is highly compressible: the compression ratios of most of the data variables can range from $30 \times$ to $1000 \times$ if compressed by using the ZLIB software \cite{zlib} with the default level 6.
HDF5 requires users to use the ``chunked'' storage layout to enable the compression, which divides a multidimensional array (referred to as a ``dataset'' in HDF5) into equal-size subarrays, each of which is compressed independently.

For parallel write operations, compressed dataset chunks are assigned to an exclusive owner process, and then partial accesses to the chunk by any other process must be transferred to the owner before compression can be applied.
Tuning chunk parameters can have a significant performance impact because the chunk size and its dimensions determine the degree of parallelism for data (de)compression and I/O.
For instance, large chunk sizes resulting in small numbers of chunks can create unbalanced workload among processes due to the chunk’s unique access ownership of write operations.
On the other hand, large numbers of chunks increase the degree of I/O parallelism but can also dramatically increase the metadata size and chunk searching time.
Therefore, chunk parameters optimized for parallel writes may often result in poor performance for parallel reads when the access patterns are orthogonal to each other.
These facts make it a challenging task to select a chunk setting that can achieve good performance for both reads and writes.

We explore a subset of HDF5 features used to implement the NOvA data concatenation, and we present our evaluation and analysis of their performance impacts.
The discussion focuses on metadata operations, raw data operations, and the end-to-end performance.
Metadata operations under this study include opening input files, creating output files, retrieving dataset metadata (dimension sizes and data types), and creating datasets.
These operations are examined under the two HDF5 metadata I/O modes, independent and collective, and different settings for metadata caching.
In particular, we study HDF5 chunk size settings and discuss their impacts on the degree of I/O parallelism and cost of the metadata operations.

Our study of raw data operations includes performance of parallel reads, collective writes, and use of small I/O buffer sizes to carry out the concatenation in multiple rounds in order to prevent running out of memory.
Our analysis is summarized at the end of the paper, giving a list of features studied and their effectiveness for certain I/O operations.
The lessons learned from this case study can provide a guideline for other scientific applications that use HDF5 as their primary I/O method.

The paper is organized as follows.
In Section \ref{sec:nova} we describe NOvA experiment detector data.
Section \ref{sec:workflow} defines the HDF5 dataset concatenation workflow, and Section \ref{sec:exp} summarizes the experimental settings.
Section \ref{sec:meta} discusses how we tuned a variety of HDF5 features related to metadata operations.
Section \ref{sec:raw} explains how we designed our parallel I/O strategy to achieve a scalable raw data I/O performance.
We report the end-to-end dataset concatenation performance with the best-tuned HDF5 feature settings and discuss the lessons we learned in Section \ref{sec:eval} and Section \ref{sec:summary}.
In Section \ref{sec:conclusion} we summarize our work and briefly discuss future
topics for research. 

%% file: 2_nova.tex
\section {NOvA Experiment Detector Data} \label{sec:nova}

\begin{table}
\scriptsize
\centering
\caption{NOvA data table organization with one entry per slice.}
\begin{tabular}{|r|r|r|r|r|r|} \hline
    Run & Subrun & Event & Sub-  & distallpngtop & ... 35 more \\
        &        &       & event &               & ... \\ \hline \hline
    433 & 61     & 6124  & 35    & nan           & \\ \hline
    433 & 61     & 6124  & 36    & -0.7401       & \\ \hline
    433 & 61     & 6124  & 37    & nan           & \\ \hline
    433 & 61     & 6125  & 1     & nan           & \\ \hline
    433 & 61     & 6125  & 2     & 423.633       & \\ \hline
    433 & 61     & 6125  & 3     & -2.8498       & \\ \hline \hline
\end{tabular}
\label{table:slice}
\end{table}

The NOvA experiment has a near detector (ND), located at the Fermilab site in Batavia, IL,
and a far detector (FD), located in Ash River, MN.
Both detectors observe neutrinos from a beam generated at Fermilab.
Approximately every 1.3~seconds, a 10~$\mu$s pulse of neutrinos is generated at the Fermilab accelerator complex and directed to the NOvA detectors.
The data collection period corresponding to one such pulse is called a \emph{spill}.
A \emph{slice} is a fixed-duration time window around a period of detector activity discovered within a spill. 
A \emph{run} is a period of data collection that represents a stable period of detector operations.
Runs have a typical temporal duration of a few hours to a maximum of 24~hours.
A \emph{subrun} is a subdivision of a run period that limits output file sizes to allow for the application of fine-grained calibration.
Subruns range from a few minutes to a maximum of 1~hour.
Data collection runs for the ND are independent of runs for the FD.
For each detector, the data from all spills in a subrun are written to a single file.


NOvA data processing proceeds through several steps, yielding increasingly detailed descriptions of the physical processes that have been observed.
In order to reduce the complexity of the management of metadata external to the files, these processing steps each process a single file (and thus a single subrun).
The final step in the processing chain includes writing out a high-level summary of each spill in forms suitable for statistical analysis of the data.
One of the output forms chosen for the data is HDF5.

In the HDF5 files, a spill is referred to as an \emph{event} and  a slice as a \emph{subevent}. 
The high-level summary data is organized in the form of different tables, suitable for the HDF5 format. 
Two examples, each depicting a different level of data, are shown in Tables~\ref{table:slice} and \ref{table:vertex}. 
In Table~\ref{table:slice}, there is one row per subevent and data showing one run, one subrun, two events, and six subevents.
In Table~\ref{table:vertex}, there are zero or more rows per subevent and one row per vertex. 
Subevents 37 and 2 (row 3 and row 5, respectively in Table~\ref{table:slice}) have no vertices since there is no entry in Table~\ref{table:vertex}. 
There are more nested levels of data, and in each corresponding table the first several columns identify runs, subruns, events, and subevents for this level.  

\begin{table}
\centering
\scriptsize
\caption{NOvA data table organization with one entry per vertex.}
\begin{tabular}{|r|r|r|r|r|r|r|} \hline
    Run & Subrun & Event & Sub-  & vtxid & npng3d & ... 6 more \\ 
        &        &       & event &       &        & ...        \\ \hline \hline
    433 & 61     & 6124  & 35    & 0     & 0      & \\ \hline
    433 & 61     & 6124  & 36    & 0     & 1      & \\ \hline
    433 & 61     & 6124  & 36    & 1     & 1      & \\ \hline
    433 & 61     & 6124  & 36    & 2     & 5      & \\ \hline
    433 & 61     & 6125  & 1     & 0     & 1      & \\ \hline
    433 & 61     & 6125  & 3     & 0     & 0      & \\ \hline \hline
\end{tabular}
\label{table:vertex}
\end{table}

When storing NOvA data in HDF5 files, each table is defined as an HDF5 group and each column as an HDF5 dataset.
All NOvA files used in the case study have the same schema, which consists of the same number of groups and datasets in each file.
All datasets are two-dimensional arrays of integer, float, or double-precision data types.
Datasets in the same group share the first dimension size.
The majority of the datasets (more than 99\%) have the second dimension of size equal to 1.
In this paper we refer to these datasets as \lq{}1D datasets\rq{} and the others as \lq{}2D datasets\rq{}.

\begin{table}
\scriptsize
\centering
\caption{Statistics of NOvA ND and FD data files.}
\begin{tabular}{|r||r|r|} \hline
    & ND data files & FD data files \\ \hline \hline
    \# of files & 165 & 6,400 \\ \hline 
    \# of groups per file & 999 & 701 \\ \hline
    \# of 1D datasets per file & 15,965 & 12,925 \\ \hline
    \# of 2D datasets per file & 8 & 6 \\ \hline
    \# of empty datasets & 13,396 & 9,374 \\ \hline \hline
    Compression & GZIP-level 6 & GZIP-level 6 \\ \hline
    Chunk size & 128-element based & 128-element based \\ \hline
    1D dsets before compr. & 97.9 GB & 413.3 GB \\ \hline
    1D dsets after compr. & 21.4 GB & 69.8 GB \\ \hline
    2D dsets before compr. & 903.2 GB & 16.6 \textbf{\textit{TB}} \\ \hline
    2D dsets after compr. & 2.1 GB & 32.1 GB \\ \hline 
    Overall before compr. & 1001.1 GB & 17.0 \textbf{\textit{TB}} \\ \hline
    Overall after compr. & 23.5 GB & 101.9 GB \\ \hline \hline
    Total file size & 35.2 GB & 212.5 GB \\ \hline
    Metadata size & 11.7 GB & 104.1 GB\\ \hline
    Raw data size & 23.5 GB & 108.1 GB \\ \hline
\end{tabular}
\label{table:statistics}
\end{table}

In the NOvA data analysis workflow, only subsets of datasets are usually selected to be analyzed together.
Once selected, datasets stored in different HDF5 files are collected into a contiguous space to be accessed as single entities.
When running data analysis in parallel, individual datasets are partitioned among processes based on their temporal IDs, such as the event or subevent datasets in the same group.
In order to achieve a good parallel efficiency and maintain data partitioning flexibility, the same datasets among all the input files are individually concatenated and saved in a new file.
This concatenation is a critical step enabling the NOvA data analysis component to search through the entire detector readouts in parallel and identify an often-small fraction that contains neutrino interactions of interest \cite{eventselection, spark}.

In this work, we study the HDF5 dataset concatenation performance using both ND and FD files.
Table \ref{table:statistics} presents the statistics we collected from ND and FD data files.
Both files contain a large number of datasets with various sizes.
The statistics can be helpful to better understand the input data and design good heuristics for parallel dataset concatenation strategy.
Each ND file in this study consists of about 16,000 datasets; each FD file consists of about 13,000 datasets.
Because there are so many datasets, a small improvement from tuning metadata parameters yields a significant result for the whole file.

One interesting fact is that about 84\% of the datasets in the NOvA files used in this work are zero in size.
The zero-size datasets represent simulation information, so detector data files will never have them. 
When this data schema was devised, it was thought that having identical schema for simulation output and detector data would be convenient. 
Although such datasets do not contain any data, we create them in the output file to provide a consistent tabular form of the data for data analysis applications.

\textbf{The impact of dataset concatenation} --
As stated earlier in this section, the size and content of files written by the
experiment are dictated by the length of the period of data collection (the
subrun) and the size of the data written during that period. The file size
limitation is imposed due to a variety of data staging mechanisms such as tape
archives, disk caching systems, as well as the grid processing farms that are
used to process the data. As a result, the raw data are stored into many small
files. The raw data go through a multi-step processing workflow before
analysis-level (``ntuple'') data files are written. These data files are
written in two formats, one of which is the HDF5 format described here. Because
the experiment's workflow is (for many reasons, unrelated to HDF5) limited to
processing a single subrun (and thus a single file) at each step, the HDF5
files written during this processing are limited to containing the data for a
single subrun. The HDF5 dataset concatenation is an offline data processing
that does not affect the data collection procedure or the ntuple-generation
workflow.

A typical analysis task is the generation of a set of histograms. With their
current system, the production of a set of histograms for one analysis is done
by running many (potentially thousands) of batch jobs on the grid, each of
which processes a few of the small ntuple files. Each of these jobs would write
out a file with the set of histograms corresponding to the subruns that job
processed. Another job that performs a reduction operation must then be run,
summing each of the histograms in the set across all the thousands of outputs
written in the first step. When aspects of the analysis are modified (which
happens frequently), the entire set of batch jobs and the reduction operation
must be repeated. With the data available in a single HDF5 file, resulting from
the concatenation described in this paper, the equivalent set of histograms can
be created by a single (MPI parallel) program run on the full data set, with no
additional reduction process needed afterward. This is a direct improvement on
the productivity of end-users.

%% file: 3_workflow.tex
\section {Dataset Concatenation Workflow} \label {sec:workflow}

We use the following notations to describe the workflow of dataset concatenation operation.
$F$ is the number of input files.
$D$ is the total number of datasets in each file.
$P$ is the number of MPI processes.
The parallel HDF5 dataset concatenation workflow is given below.

\begin {enumerate}
    \item {Evenly distribute $F$ input files to $P$ processes.}
    \item {Collect and aggregate dimension sizes of $D$ datasets from all input files.}
    \item {Create a new shared output file.}
    \item {Create $D$ datasets of aggregated sizes in the output file.}
    \item {For each dataset, read the dataset from all assigned input files to a memory buffer by appending one after another, and then write the concatenated buffer to the output file.}
\end {enumerate}

To balance the workload, we first evenly distribute the given $F$ input files to $P$ processes such that each process is assigned with $\frac{F}{P}$ different files.
Then, each process opens the assigned files.
In step 2, each process collects the data types and array sizes of individual datasets from the assigned files.
The locally collected array sizes are aggregated among all $P$ processes by using \texttt{MPI\_Allreduce}.
In step 3, all processes collectively create a new output file.
In step 4, all processes collectively create $D$ datasets with the aggregated sizes.
In step 5, for each dataset all the processes independently read the dataset from the assigned $F/P$ files into a memory buffer by appending one after another and then collectively write the concatenated buffer to the shared output file.
The output file uses the same data object schema as the input files use, in other words, the same groups, datasets, and their memberships.

%% file: 4_exp_settings.tex
\section {Experimental Settings} \label {sec:exp}

All our experiments are conducted on Cori, a Cray XC40 supercomputer at NERSC.
We ran only on Haswell nodes because they have larger memory and faster I/O speed than the KNL nodes have.
Each Haswell node has 128~GiB DDR4 2133 MHz memory, compared with 96~GiB on KNL.
Each Haswell node has two sockets of Intel Xeon E5-2698v3 CPUs with 16 cores each.
In our experiments, all the input files and output files are stored on a disk-based Lustre parallel file system, using a file stripe size of 1~MB and a file stripe count 128.
The parallel HDF5 library used in our experiments is version 1.10.5.

%% file: 5_meta.tex
\section {Metadata I/O} \label{sec:meta}

Step 2 of  the workflow described in Section \ref{sec:workflow} performs metadata read operations for collecting data type and array sizes of all the datasets.
The array size metadata is then aggregated into a global size, which is then used to create new datasets in the output file in steps 3 and 4.
Reading and writing chunked and compressed raw data in step 5 also involve metadata operations that traverse the internal B-trees.
A B-tree is a self-balancing tree data structure adopted by HDF5 for fast data object lookup, such as searching for object names and locations of data chunks.
For each I/O request to chunked datasets, HDF5 traverses the relevant B-trees to find all the chunks whose space intersects with the request.
Given $13K \sim 16K$ datasets in each input file and hundreds to thousands of files, concatenating all individual datasets is expected to be metadata operational expensive.
In this section, we focus on studying various HDF5 features and analyzing their impact on the metadata operation performance.

\subsection {Reading Metadata from Input Files} \label{sec:input}

In order to concatenate individual datasets, their data types and array sizes must be first collected and aggregated across all input files.
In step 2 of the workflow, each MPI process reads such metadata from the assigned files.
Since each process is assigned a distinct subset of the input files, metadata can be read independently.
Once the metadata is collected, the local array sizes are summed among all the processes with an MPI collective communication call to {\tt MPI\_Allreduce}.
The aggregated array sizes will be used to define new datasets in step 4.
In HDF5, a file can be opened in either POSIX or MPI I/O mode.
Using POSIX I/O mode is equivalent to using the MPI I/O mode with the communicator set to {\tt MPI\_COMM\_SELF}.
If the MPI I/O mode is used, a negligible additional cost over the POSIX mode is expected, due to argument sanity checks performed in the MPI library.

\textbf{In-memory I/O vs. On-the-fly I/O} --
HDF5 adopts a flexible file format that allows metadata and raw data of individual data objects to be stored separately in locations almost anywhere in the file.
Thus, collecting metadata in step 2 may result in read operations on noncontiguous file regions.
Given the large number of datasets in NOvA files, step 2 can become expensive if the number of noncontiguous metadata file blocks is high.
To mitigate the I/O cost, HDF5 provides an in-memory I/O feature that can load the entire file into the memory at file open time, so the successive requests to the file can be fulfilled through memory copy operations.
The HDF5 in-memory I/O feature is enabled by specifying the \textit{core} file driver with a call to {\tt H5Pset\_fapl\_core} API, and the entire file is loaded into an internal buffer when opening the file.
Note that in-memory I/O is currently supported only for POSIX I/O mode.

\begin{table}
\scriptsize
\centering
\caption{Metadata collection time (sec) for the 165 ND files.
    Up to 42 compute nodes and 165  MPI processes were used in our evaluation, and in each case the MPI processes are evenly assigned to the compute nodes.
    Compared with the on-the-fly I/O method, the in-memory I/O method shows significantly lower costs in reading the metadata.
}
\begin{tabular}{|r||r|r|r|r|r|r|r|} \hline
     \# of processes & 3 & 6 & 11 & 21 & 42 & 83 & 165 \\ \hline
     \# of nodes & 1 & 2 & 3 & 6 & 11 & 21 & 42 \\ \hline
     On-the-fly I/O & 4099 & 2188 & 1122 & 663 & 304 & 131 & 64 \\ \hline
     In-memory I/O & 61.9 & 26.8 & 16.3 & 8.9 & 4.9 & 3.6 & 1.5 \\ \hline
\end{tabular}
\label{table:meta_nd}
\end{table}

\begin{table}
\scriptsize
\centering
\caption{Metadata collection time (sec) for the 6,400 FD files.
    Up to 400 compute nodes and 1,600  MPI processes were used in our evaluation, and in each case the MPI processes were evenly assigned to the compute nodes.
    Compared with the on-the-fly I/O method, the in-memory I/O method shows significantly lower costs in reading the metadata.
}
\begin{tabular}{|r||r|r|r|r|r|} \hline
     \# of processes & 100 & 200 & 400 & 800 & 1600 \\ \hline
     \# of nodes & 25 & 50 & 100 & 200 & 400 \\ \hline
     On-the-fly I/O & 355.8 & 180.2 & 93.4 & 55.0 & 37.1 \\ \hline
     In-memory I/O & 42.1 & 21.3 & 13.9 & 8.7 & 5.7 \\ \hline
\end{tabular}
\label{table:meta_fd}
\end{table}

Tables \ref{table:meta_nd} and \ref{table:meta_fd} compare the metadata read performance between in-memory and on-the-fly I/O methods for the ND and FD files, respectively.
We used up to 165 MPI processes on 42 compute nodes for the 165 ND files and up to 1,600 processes on 400 nodes for the 6,400 FD files.
The timing results show that the in-memory I/O significantly improves the metadata read performance.
Such a big gap is realized by checking the metadata locations for individual datasets in the input files, revealing a high number of noncontiguous file regions.

Note that when in-memory I/O is enabled, the raw data is also loaded into memory, increasing the memory footprints.
This in-memory approach may become infeasible for files containing very large raw data.
In our concatenation case, when the number of files assigned to a process increases, the increased memory footprint required by each process can limit the number of processes running on each compute node.
Many HDF5 features also consume memory, such as a file's metadata cache, raw data chunk cache, and internal memory buffers for data compression and decompression.
As shown in Tables \ref{table:meta_nd} and \ref{table:meta_fd}, we were able to run up to 4 processes on each compute node without encountering the out-of-memory error.
More discussion on the memory footprint analysis will be presented in Section \ref{sec:buffer}.

\subsection{New Dataset Creation} \label{sec:create}

In step 4, new datasets of concatenated sizes are created in the output file based on the metadata collected in step 2.
We consider two possible design options for parallel dataset creation.
One is to let a single process create all the datasets first, followed by having all the processes open the created datasets collectively.
In this approach, the creating process can open the output file in POSIX I/O mode and create all the datasets without MPI communication cost.
The other option is to have all the processes open the output file in MPI I/O mode and collectively create all the datasets.

\begin{figure}
\centering
\includegraphics[width=\columnwidth]{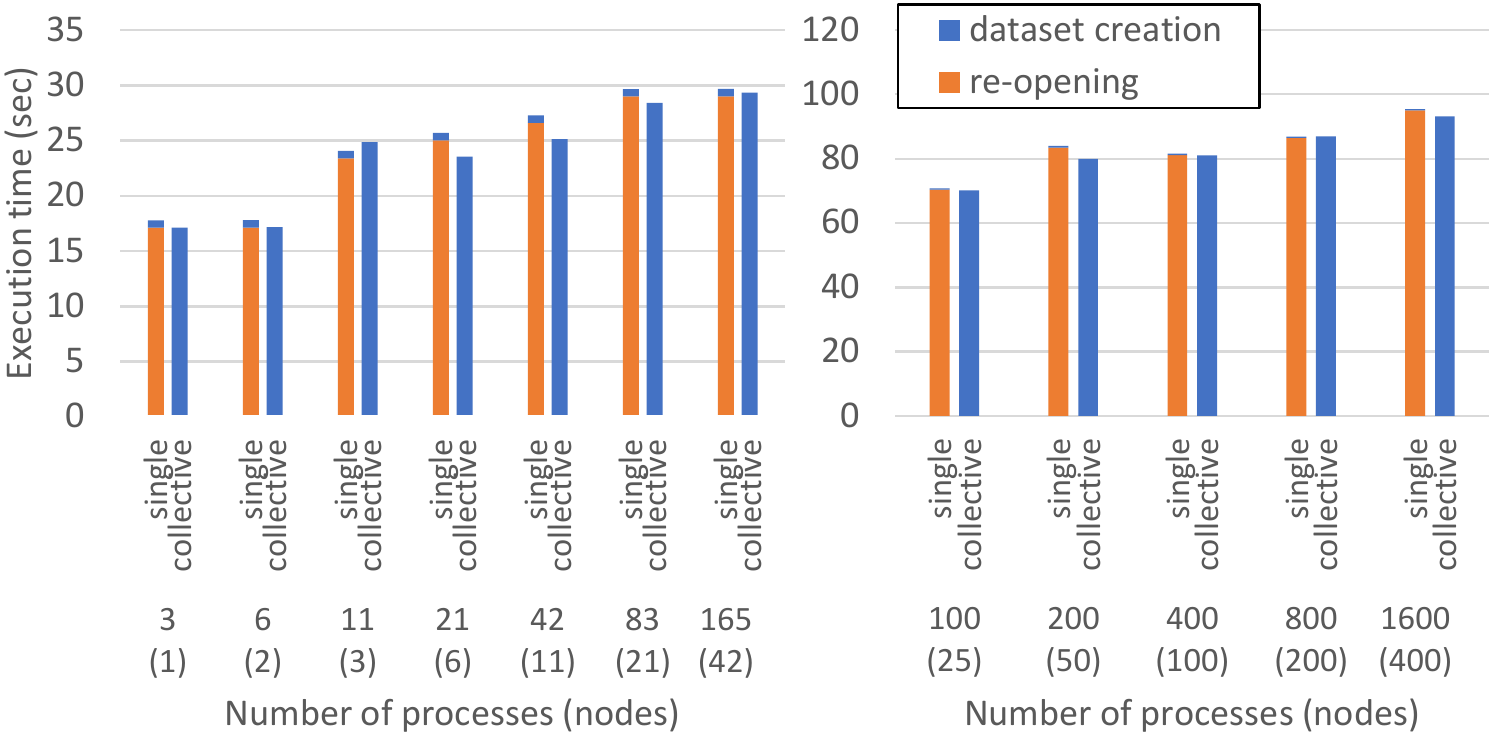}
\caption{
    Timing breakdowns for creating all the datasets when concatenating 165 ND files (left) and 6,400 FD files (right).
    Bars labeled with \lq{}single\rq{} and \lq{}collective\rq{} represent timings of the single-process creation method and collective creation method, respectively.
}
\label{nd_create}
\end{figure}

Figure \ref{nd_create} presents timing breakdowns for creating all the datasets using these two options.
The left chart is for the 165 ND files and the right for the 6,400 FD files.
We observed that the single-process creation option spends much less time on {\tt H5Dcreate} than the collective creation option does. 
However, it takes more time to reopen the datasets in parallel I/O mode.
The end-to-end time for both options shows a similar performance.
This behavior can be explained by the HDF5 design of the dataset fill mode.

In the HDF5 implementation, a compression-enabled new dataset must be first filled with either a predefined or a user-supplied fill value.
When the file is created in nonparallel mode, the HDF5 policy is to delay the file space allocation for new datasets to when they need to be written.
Since the first creation option does not write any raw data, the new datasets are not filled in the file.
Later, when opening the datasets in parallel, HDF5 detects the file space yet to be allocated and starts filling the datasets in the file.
This approach explains why the first option has a shorter creation time and a longer opening time.
On the other hand, when the file is created in parallel mode, the HDF5 policy is to allocate the file space for a new dataset and fill it immediately during the call of {\tt H5Dcreate}.
The fact that the same data-filling operation is required by HDF5 at either {\tt H5Dcreate} or {\tt H5Dopen} explains the similar timings observed for the two dataset creation options.

To mitigate the data-filling cost, HDF5 is considering adjusting the implementation by moving the filling operation to the first call of {\tt H5Dwrite}, when the write patterns to individual chunks are known.
For instance, if a chunk is entirely written, then data filling for that chunk can be skipped.
We expect such an optimization will significantly reduce the dataset creation cost.

\textbf{Independent vs. collective metadata I/O mode} --
HDF5 provides two I/O modes for metadata operations: \textit{independent} and \textit{collective}.
The default setting is independent mode.
Collective metadata operations can be enabled through APIs {\tt H5Pset\_coll\_metadata\_write} and {\tt H5Pset\_all\_coll\_metadata\_ops}.
The former is for write operations and the latter for reads.
To achieve a better parallel performance, the HDF Group suggests collective metadata I/O mode, especially when the metadata size is large \cite{meta_coll}.
In our experiments we observed no noticeable difference between the two I/O modes for dataset creation time (less than 1 second), but a large disparity at file close time, {\tt H5Fclose}.

HDF5 performs internal metadata caching and can automatically adjust the cache buffer size to accommodate more metadata \cite{meta_coll,meta_coll2}.
Cached metadata is flushed to the file system when the accumulated size increases beyond a defined threshold or at the file close time.
In our case, even when creating 16K datasets, the accumulated metadata size appears to be small enough to be kept in the cache without triggering a flush during the dataset creation loop.
On the  other hand, we observe 4.97 seconds spent on {\tt H5Fclose} for independent mode and 0.32 seconds for collective mode, when running 165 processes on the ND files.

Such behavior can be explained by how the metadata flushing mechanism is implemented differently for independent and collective modes.
For independent mode, each metadata block is written to the file by a call to the independent MPI file write function.
For collective mode, HDF5 uses an MPI derived data type to describe the memory layout of multiple (noncontiguous) metadata blocks, so they can be written in a single call to the collective MPI file write function.
A performance benchmark for metadata caching can be found in \cite{meta_coll2}.

\subsection {Chunking and Compression Strategy}\label{sec:chunk}

HDF5 datasets must use the chunked layout to enable compression of data elements.
Parameter tuning for both chunk size and compression can have a significant impact on the parallel I/O performance scalability.
Chunk size is set by using the API {\tt H5Pset\_chunk} and compression level {\tt H5Pset\_deflate}.
The current HDF5 implementation supports only the collective I/O mode for writing compressed datasets in parallel.

\textbf{Chunk ownership} --
For parallel write operations, HDF5 first assigns each  chunk of a dataset to a unique \lq{}owner\rq{} process.
The chunk owner is responsible for collecting write requests from all other processes, compressing the chunk, and writing to the file.
The assignment policy for chunk ownership is to assign the chunk owner to the process whose write request covers the largest part of the chunk.
If a chunk is written by multiple processes in parallel, data for each chunk is first transferred from non-owners to the owner.
Once all data transfers complete, chunk owners compress the chunks using an external compression library, such as ZLIB (default) \cite{zlib} or SZIP \cite{szip}.
Because of the uniqueness of chunk ownership, data compression can be performed concurrently among all owners.
The compressed data chunks are then collectively written into the output file.

\begin{figure*}
\centering
\includegraphics[width=2\columnwidth]{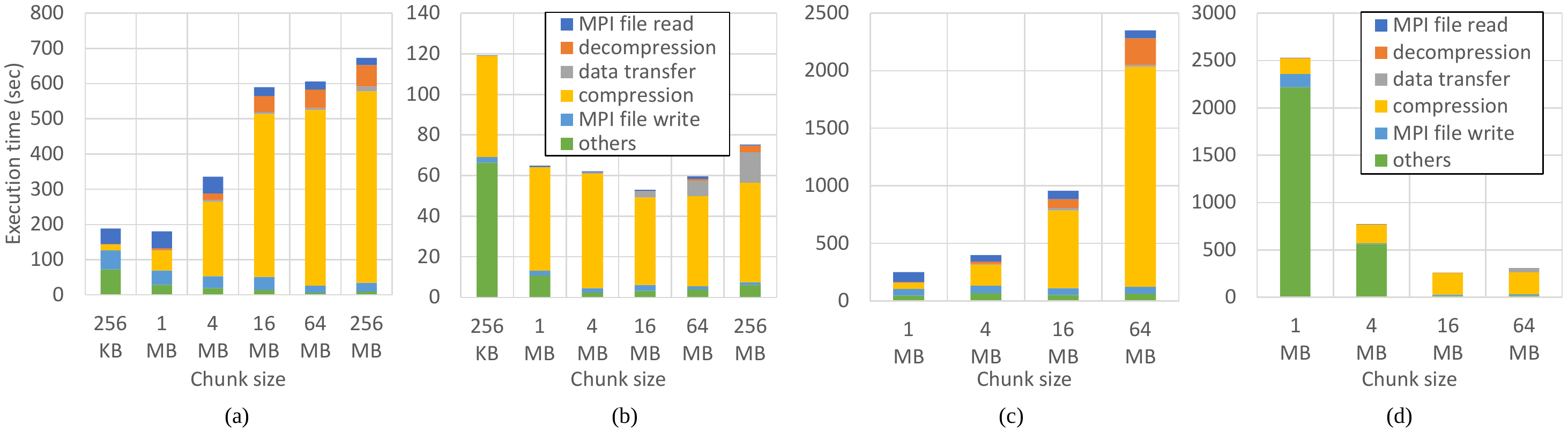}
\caption{
    Timing breakdown for writing 1D ND datasets (a), 2D ND datasets (b), 1D FD datasets (c), and 2D FD datasets (d) with varying chunk sizes.
    For ND files, we ran 165 processes on 42 nodes.
    For FD files, we ran 800 processes on 200 nodes.
    Since HDF5 performs \lq{}read-modify-write\rq{} for chunks that are not written  completely by their owners, \lq{}MPI file read\rq{} represents the read time for such I/O operations.
}
\label{chunk}
\end{figure*}

{\bf I/O parallelism} --
The HDF5 dataset write API routine {\tt H5Dwrite} allows access to a single dataset at a time.
Because of the unique chunk ownership policy, the degree of write parallelism to a dataset is determined by the number of its chunks.
According to the HDF5 User Guide, the chunk size should be sufficiently small so that there are enough chunks to keep all the processes busy in performing compression and I/O.
In particular, if there are fewer chunks than the number of processes, then only a subset of processes will compress and write the chunks while others stay idle.
On the other hand, chunk size should be big enough to obtain good compression ratios.

Additionally, too many chunks due to small chunk sizes can increase the metadata operation cost.
Currently, tuning the chunk size can be done only by the user, because of the wide variety of access patterns involved.
Factors that should be considered include dataset size, data partitioning  pattern, number of application processes, and compression level.
By default, chunks of a dataset are indexed with a B-tree data structure for fast lookup.
Deep B-trees are more expensive to traverse; however, B-trees with larger nodes and a shallower depth use more memory for each node.
In HDF5, the size of a B-tree node can be adjusted by setting the rank (called ``k'') of nodes with the {\tt H5Pset\_istore\_k} API.
HDF5 uses $2k$ as the maximum number of entries before splitting a B-tree node.
The default value of $k$ for chunk index B-tree nodes is $32$, and the maximum allowed value is $32,768$.
In this case study, we used the default value only.

Figure \ref{chunk} presents the timing breakdown for writing 1D and 2D datasets when concatenating 165 ND files (a and b) and 6,400 FD files (c and d).
For ND files, we measured the write timings using chunk sizes between 256 KB and 256 MB.
Note that HDF5 limits each dataset to at most $2^{32}$ chunks.
For FD files, the number of chunks for the largest dataset can go over the limit if the chunk size is less than 1 MB.
Therefore, we set the chunk sizes to between 1 MB and 64 MB for FD files.
In HDF5, a \textit{read-modify-write} operation is performed for chunks that are not completely written by their owners.
This design choice is to consider the possibility of a chunk that is partially written.
\lq{}MPI file read\rq{} indicates the time spent on such read operations.
\lq{}Others\rq{} is calculated by subtracting all the labeled times from the end-to-end time of {\tt H5Dwrite}.
It mainly represents the metadata operation time.

{\bf Impact  of chunk size to the costs of compression and data transfer} --
For the 1D datasets as shown in Figures \ref{chunk} (a) and (c), setting the chunk size to larger than 1 MB increases the end-to-end write time.
This is due to the increasing compression time.
For example, there are 1,458 ND datasets of size smaller than 128 MB after the concatenation.
Setting chunk size larger than 1 MB results in the number of chunks being smaller than the number of processes, which makes some processes owning no chunks.
For the 2D datasets shown in Figures \ref{chunk} (b) and (d), since their sizes are much bigger than those of the 1D datasets, setting chunk sizes smaller than 4 MB increases the B-tree sizes and cost of metadata operation.
The chunk sizes larger than 16 MB also degrade the write performance  because of the increased time spent on data transferring from non-owners to the chunk owners.

{\bf Cost of read-modify-write} --
We observe that 1D datasets have a longer \lq{}MPI file read\rq{} time than 2D datasets.
This is because the number of write requests from each process to the whole chunks is small in 1D datasets and large in 2D datasets.
For both ND and FD files, the majority of 1D datasets are small in each input file.
Since the dataset sizes in each input file represent the write amount in the collective write operations, write requests for 1D datasets mostly cover partial chunks.
For 2D datasets, write requests from each process are much larger, resulting in most of them covering the whole chunks.
We also observe a small amount of time spent on \lq{}MPI file write\rq{}.
This is because the NOvA data is highly compressible, which results in a small write amount and thus contributes insignificantly to the end-to-end write time.

\textbf{Compact data layout for zero-size datasets} --
NOvA files contain a large number of zero-size datasets.
For example, each ND data file has 15,973 datasets in total, and 13,392 of them have their first dimensions of size zero.
The concatenated file uses the same data object schema as the input files do; that is, the same names and hierarchies of groups and datasets as the concatenated file will also be used by other HEP data analysis applications.
HDF5 supports three dataset layouts---\textit{contiguous}, \textit{chunked}, and \textit{compact}---with {\it contiguous} being the default.
Datasets in contiguous layout store their raw data in a single contiguous block, at an arbitrary offset in the file.
The raw data of a dataset in chunked layout is split into multiple chunks, which are stored at arbitrary offsets in a file.
The file locations storing chunks may not even be in an increasing file offset order, since HDF5 optimizes the allocations of chunks to prevent any wasted space \cite{hdf5}.
Chunked layout allows applications to define datasets with extendible dimensions and is also required for compressing raw data.
Compact layout stores the raw data of a dataset within its object header but is available only  when the raw data size is smaller than 64~KB, the maximum HDF5 object header size.

\begin{table}
\scriptsize
\centering
\caption{
    Comparison of zero-size dataset creation time among 3 data layout settings, when concatenating 165 ND files running 165 processes on 42 nodes.
}
\begin{tabular}{|r||r|r|r|} \hline
    Layout & \textit{contiguous} & \textit{chunked} & \textit{compact} \\ \hline \hline
    Dataset creation time (sec) & 30.9 & 39.2 & 30.9 \\ \hline
    Metadata size & 29.3 MB & 39.2 MB & 29.3 MB \\ \hline
    Raw data size & 15.0 GB & 15.1 GB & 15.0 GB \\ \hline
    Overall file size & 15.0 GB & 15.2 GB & 15.0 GB \\ \hline
\end{tabular}
\label{table:layout_nd}
\end{table}

\begin{table}
\scriptsize
\centering
\caption{
        Comparison of zero-size dataset creation time among 3 data layout settings, when concatenating 6,400 FD files running 800 processes on 200 nodes.
}
\begin{tabular}{|r||r|r|r|} \hline
    Layout & \textit{contiguous} & \textit{chunked} & \textit{compact} \\ \hline \hline
    Dataset creation time (sec) & 64.1 & 73.8 & 64.0 \\ \hline
    Metadata size & 93.8 MB & 119.5 MB & 93.8 MB \\ \hline
    Raw data size & 77.4 GB & 77.4 GB & 77.4 GB \\ \hline
    Overall file size & 77.6 GB & 77.6 GB & 77.6 GB \\ \hline
\end{tabular}
\label{table:layout_fd}
\end{table}

For zero-size datasets, we evaluated the dataset creation performance of using three data layouts and studied their impact on the concatenated output file size.
Tables \ref{table:layout_nd} and \ref{table:layout_fd} show the dataset creation time and data sizes in the concatenated output file, respectively.
Using chunked layout for zero-size datasets results in a larger metadata size and a slower dataset creation time than those of the other two layouts.
The reason  is that  HDF5 still generates the chunking metadata such as B-trees, regardless of the data size.
Because the majority of 16K datasets are of size zero, the accumulated chunking metadata amount and time for creating the metadata can become significant.
HDF5 developers are aware of this behavior and currently are developing a fix to avoid B-tree allocation for zero-size datasets; the fix is expected to be available in the next release (version 1.12.1).
When using the contiguous layout, HDF5 produces the output file of size exactly the same as the compact layout, since both layouts neither create B-tree metadata nor occupy raw data space.
With the same metadata sizes and operations, these two layouts also exhibit similar dataset creation costs.

\begin{table}
\scriptsize
\centering
\caption{
    Comparison of parallel write performance and output file size when using different compression levels to concatenate 165 ND files running 165 processes on 42 nodes.
}
\begin{tabular}{|r||r|r|r|r|} \hline
  Compression level & 2 & 4 & 6 & 8 \\ \hline \hline
  Compression time (sec) & 89.4 & 92.6 & 100.0 & 181.2 \\ \hline
  MPI File write time (sec) & 31.6 & 32.7 & 35.3 & 26.9 \\ \hline
  Output file size (GB) & 19.0 & 15.2 & 15.0 & 14.9 \\ \hline
\end{tabular}
\label{table:comp_nd}
\end{table}

\textbf{Data compression level} --
All NOvA files used ZLIB compression of level 6 when they were generated.
ZLIB supports 10 levels ($0 \sim 9$) of data compression.
The compression level can be set by using {\tt H5Pset\_deflate} API.
With a lower compression level, a faster compression time is expected but with a lower compression ratio.
In this paper, we measure the compression ratio as the ratio of the uncompressed size over the compressed size.
Tables \ref{table:comp_nd} and \ref{table:comp_fd} show the write performance and output file sizes for 165 ND files and 6,400 FD files, respectively.
The concatenation ran 165 MPI processes on 42 nodes for ND files and 800 MPI processes on 200 nodes for FD files.
For ND files, we observe that level 2 achieves the shortest MPI file write time but produces the largest file size.
As the compression level goes up, the compression time increases, and both write time file sizes are reduced.
When using level 8, the file size is not much smaller than using level 6, but the compression time significantly increases.
For FD files, level 2 results in the longest write time due to the poor compression ratio.
Similar to the ND case, level 8 suffers from a longest compression but produces the smallest output file.

\begin{table}
\scriptsize
\centering
\caption{
    Comparison of parallel write performance and output file size when using different compression levels to concatenate 6,400 FD files running 800 processes on 200 nodes.
}
\begin{tabular}{|r||r|r|r|r|} \hline
  Compression level & 2 & 4 & 6 & 8 \\ \hline \hline
  Compression time (sec) & 95.3 & 249.4 & 278.6 & 364.0 \\ \hline
  MPI File write time (sec) & 22.6 & 13.9 & 8.3 & 8.4 \\ \hline
  Output file size (GB) & 142.0 & 78.8 & 77.6 & 75.8 \\ \hline
\end{tabular}
\label{table:comp_fd}
\end{table}

\subsection {Metadata Caching}

Given the fact that there are about 16K datasets to be created in our case study, the number of metadata creation operations is expected to be high.
Caching metadata in memory and later flushing the metadata in bigger, aggregated write requests appears to be a good strategy to achieve good performance.
HDF5 enables metadata caching by default and uses an initial metadata cache size of 2 MB, which is automatically adjusted based on the cache hit rate.
The metadata cache is implemented with a hash table that indexes a pool of varying-size metadata entries.

To evaluate its performance impact, we tested and set the initial cache size to 128 MB, the maximum cache size allowed by the hash table size in HDF5.
We also disabled the automatic size adjustment feature so that the metadata cache size is fixed to 128 MB.
With this configuration, we expected to minimize the cost of memory operations to expand the cache size and data movement between buffers during dataset creation and the collective write operations.
Note that HDF5 also allows the hash table size to be increased, to use a metadata cache larger than 128 MB.
However, in our case study we found that using a larger cache size does not further improve the performance, particularly after the proper chunk size, for example, 1 MB based, is used to effectively reduce the size of B-trees for dataset chunk indices.

Tables \ref{table:metacache_nd} and \ref{table:metacache_fd} present the performance gains of using this metadata cache setting over the default setting for ND and FD files, respectively.
For ND files, the case shown in the table is for concatenating 165 ND files using 165 processes on 42 nodes.
For FD files, the case is for concatenating 6,400 ND files using 800 processes on 200 nodes.
The timings of dataset creation and parallel writes are reduced when the metadata cache size increases.
In particular, 1D dataset write performance is significantly improved because a large portion of the write time is taken by the metadata operations.
With a large metadata cache size, the frequency of metadata eviction can be reduced.
In addition, when the metadata cache is sufficiently large, the B-tree metadata created and cached in step 4 can be reused in step 5.
Thus, increasing metadata cache size effectively improves the performance for the dataset  concatenation in our case study.

\begin{table}
\scriptsize
\centering
\caption{
    Performance comparison (seconds) between the default metadata cache size and larger cache size setting for concatenating 165 ND files running 165 processes on 42 nodes.
    The default setting has an initial cache size of 2~MB and is automatically adjusted based on the cache hit rate.
    Our setting is with an initial size of 128~MB and disables automatic cache size adjustment.
}
\begin{tabular}{|r||r|r|r|} \hline
    Steps & Default & Our setting & Performance gain \\ \hline \hline
    Metadata collection & 1.5 & 1.5 & - \\ \hline
    Dataset creation & 30.7 & 29.3 & 4.5\% \\ \hline
    1D dsets Read & 15.5 & 15.9 & - \\ \hline
    1D dsets Write & 255.2 & 167.2 & 34.5\% \\ \hline
    2D dsets Read & 17.2 & 17.3 & - \\ \hline
    2D dsets Write & 54.1 & 48.2 & 10.9\% \\ \hline
    Overall & 374.3 & 279.4 & 25.3\% \\ \hline
\end{tabular}
\label{table:metacache_nd}
\end{table}

\subsection{Adjustment of Metadata Block Size}

In HDF5, metadata blocks are contiguous regions in a file that store the metadata, and HDF5 attempts to aggregate many metadata entries into each block.
Small metadata block sizes increase the flexibility for dynamically adding new data objects, since they provide HDF5 a better chance of finding a free location among spaces occupied by existing data objects to accommodate the new metadata block.
However, this flexibility can also result in many metadata blocks spread out in the file in noncontiguous locations.
In HDF5 the metadata block size is 2~KB by default, which is adjustable through a call to the {\tt H5Pset\_meta\_block\_size} API.
Note that if the metadata cache size is smaller than the metadata block size, a metadata block can be written in multiple rounds.

Our study shows that increasing the metadata block size for input files does improve the metadata read performance in step 2.
In step 2, we measured the time spent on {\tt H5Ovisit} API that traverses over all the objects in the input files.
The callback function used for {\tt H5Ovisit} reads the dataset object headers to retrieve the information about dimension sizes and data types.
If such metadata is stored contiguously, the {\tt H5Ovisit} performance can be improved.
With the default block size, collecting the metadata from 165 ND files using 165 processes on 42 nodes takes about 11~seconds.
When the metadata block size is adjusted to 32~MB, the metadata collection time is reduced to about 4~seconds.
Note that adjusting a file's metadata block size can be done in a postprocessing step with the HDF5 utility program \textit{h5repack} and the {\it -M} command line option.
This evaluation indicates that the metadata block size should be appropriately increased if the number of objects is large and visiting all of them is planned.

\begin{table}
\scriptsize
\centering
\caption{
    Performance comparison (seconds) between the default metadata cache setting and larger cache size setting for concatenating 6,400 FD files running 800 processes on 200 nodes.
    The default setting has an initial cache size of 2~MB and is automatically adjusted based on the cache hit rate.
    Our setting is with an initial size of 128~MB and disables automatic cache size adjustment.
}
\begin{tabular}{|r||r|r|r|} \hline
    Steps & Default & Our setting & Performance gain \\ \hline \hline
    Metadata collection & 6.8 & 6.9 & - \\ \hline
    Dataset creation & 69.7 & 64.0 & 8.2\% \\ \hline
    1D dsets Read & 11.6 & 10.9 & - \\ \hline
    1D dsets Write & 444.6 & 281.9 & 36.6\% \\ \hline
    2D dsets Read & 17.2 & 17.3 & - \\ \hline
    2D dsets Write & 412.4 & 272.0 & 34.0\% \\ \hline
    Overall & 1000.9 & 690.7 & 31.0\% \\ \hline
\end{tabular}
\label{table:metacache_fd}
\end{table}

%% file: 6_raw.tex
\section {Raw Data I/O} \label{sec:raw}

\begin{figure*}
\centering
\includegraphics[width=2\columnwidth]{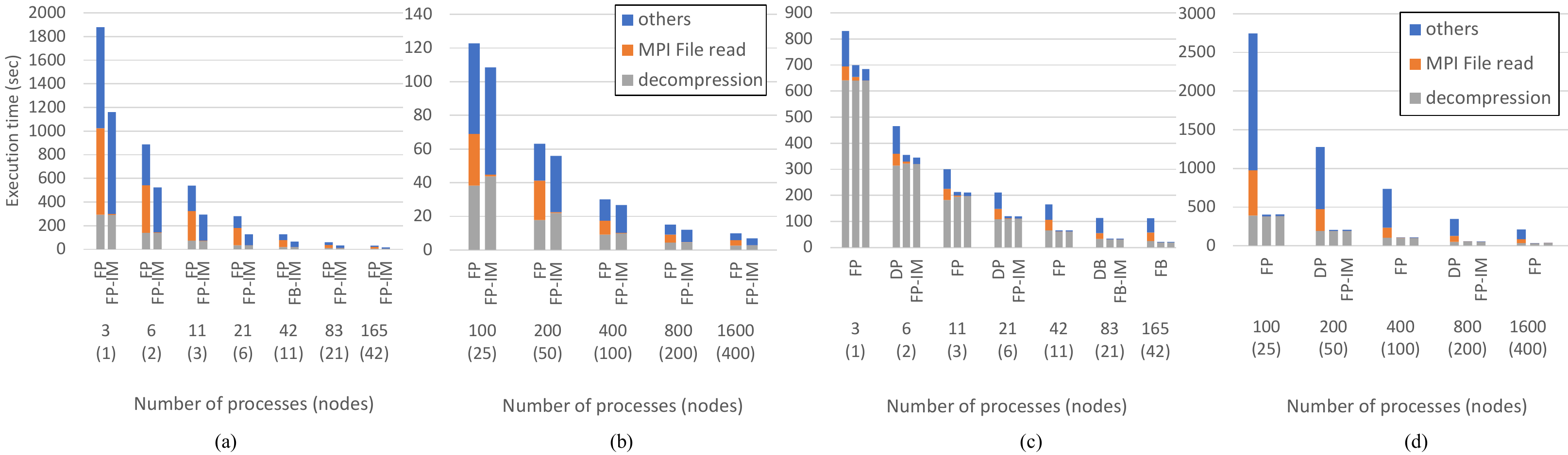}
\caption{
    Timing breakdowns for reading 1D datasets from 165 ND files (a), 6,400 FD files (b), 2D datasets from 165 ND files (c), and 6,400 FD files (d).
    We compare the performance between dataset-based partitioning (DP) and file-based partitioning (FP).
    FP-IM represents the FP with in-memory I/O.
    \lq{}others\rq{} includes the metadata operation time.
}
\label{read}
\end{figure*}

HDF5 supports parallel I/O for both shared and separate files.
For parallel I/O to separate files where each process accesses a unique set of files, one can use the default POSIX I/O file access property when opening the file.
For parallel I/O to shared files, all processes must use an MPI communicator in the file access property when opening the file with {\tt H5Fopen}.
As with collective and independent I/O modes available for metadata operations, HDF5 also allows users to select the two I/O modes for raw data operations.
Furthermore, HDF5 allows different I/O modes for individual datasets.
Specifically, 
the collective or independent data transfer property can be used when performing I/O to a dataset.
For parallel read operations in this case study, we implement two strategies, using separate- and shared-file reads.
For parallel write operations, we use the collective mode and shared-file option only.

\subsection {Parallel Read} \label{sec:read}

One possible parallel read strategy is to open the input files in MPI I/O mode and collectively read every dataset in each file.
In this approach, the number of collective reads is equal to the number of input files multiplied by the number of datasets, namely, $|F||D|$.
For large datasets, this read strategy should perform reasonably well.
For small datasets,  however, the collective read operations can underperform with the available I/O bandwidth.
NOvA files have many datasets that are smaller than 1 MB in each file.
Based on this use case, we implement two read strategies.
One reads individual 2D datasets collectively, and the other reads them independently.
Both strategies write the 2D datasets collectively to the output file.
Since the 1D datasets in NOvA files are relatively small, both strategies let each process read the 1D datasets independently from the disjointly assigned input files and write them collectively to the output file.
We refer to the first strategy as {\it dataset-based partitioning} and the second as {\it file-based partitioning}.
The dataset-based partitioning method requires all processes to open all input files so each of 2D datasets can be read collectively.
The file-based partitioning only requires each process to open the assigned input files.

File-based partitioning has two advantages.
First, because each process accesses a distinct subset of the input files, the datasets can be independently read using POSIX I/O operations, avoiding synchronization delays.
Second, the number of read operations is $\frac{FD}{P}$; therefore, 
as the number of processes $P$ increases, the datasets are read from more files at once, and thus the overall number of read operations is reduced.
One potential drawback of this approach is the unbalanced workload among the processes when datasets in input files are very different in size.
If a dataset has a high variance of size among the input files, some processes will handle more data than the other processes, causing poor scaling efficiency.

We compare the read performance between the two parallel read strategies.
Figure \ref{read} presents timing breakdowns for 1D datasets in 165 ND files (a), 1D datasets in 6,400 FD files (b), 2D datasets in ND files (c), and 2D datasets in FD files (d).
In the charts, \textit{DP} and \textit{FP} represent dataset-based partitioning and file-based partitioning, respectively.
\textit{FP-IM} is the file-based partitioning with in-memory I/O.
Each bar consists of the three workloads: data decompression, MPI File read, and others (metadata operations).

We observe that the in-memory I/O improves the read performance for both 1D and 2D datasets.
When in-memory I/O is enabled, the entire input files are preloaded into memory space when collecting the metadata in step 2.
Thus, the I/O operations become memory operations, and the \lq{}MPI File read\rq{} timing is reduced.
We also see that for 2D datasets file-based partitioning outperforms the dataset-based partitioning, as shown in Figures \ref{read}(c) and \ref{read}(d).
The performance difference between the two strategies comes mainly from the number of I/O operations.
The dataset-based partitioning performs a collective read for every dataset in all the files, whereas the file-based partitioning performs independent reads on the assigned files only.

Therefore, given the same data size, the file-based partitioning is expected to provide a shorter I/O time.
Additionally, the dataset-based partitioning suffers from expensive metadata operations.
When the dataset-based partitioning is used, for every dataset in each input file, all the processes calculate the intersection between the requested file regions and the dataset chunks and then perform an interprocess communication to synchronize the information. 
In contrast, the file-based partitioning allows all the processes to access only the assigned files in POSIX mode and has much cheaper metadata operations.

In our experiments, we set the stripe count of the Lustre parallel file system to 128.
When using the dataset-based partitioning, the collective read operations can take advantage of the distributed file system storage.
In contrast, when using file-based partitioning, every process independently reads the assigned files, and many processes can try to access the same storage server, potentially causing congestion.
We can expect a better read performance of file-based partitioning when the stripe count is reduced.

\subsection {Parallel Write} \label{sec:write}

\begin{figure*}
\centering
\includegraphics[width=2\columnwidth]{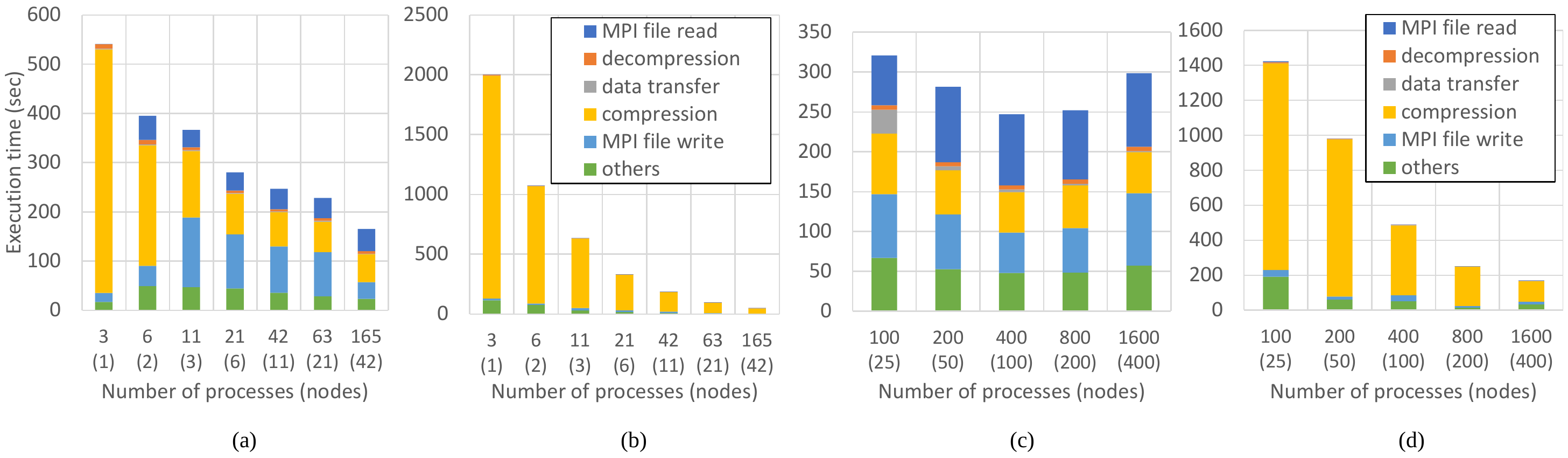}
\caption{
    Timing breakdowns for writing 1D datasets of 165 ND files (a), 2D datasets of 165 ND files (b), 1D datasets of 6,400 FD files (c), and 2D datasets of 6,400 FD files (d).
    For ND and FD files, the datasets are concatenated using up to 165 processes on 42 nodes and 1,600 processes on 400 nodes, respectively.
}
\label{write}
\end{figure*}

For both 1D and 2D datasets, once all the processes read a dataset from their assigned input files, they collectively write the concatenated dataset into the output file.
HDF5 allows applications to read or write only a single dataset at a time, so unbalanced workloads among processes may appear in our case.
When writing compressed datasets, only the ranks that own chunks perform data compression and have data to write.
For small datasets, when the number of chunk owners is less than the number of MPI processes, processes that do not own any chunks will participate the collective writes but have no data to write.
The more processes sitting idle, the worse the scalability for parallel write performance.
In our case study, there are many small datasets in NOvA files.
For ND files there are 2,573 1D datasets and for FD files there are 3,552 1D datasets, most of them being only a few kilobytes per file.

Figure \ref{write} presents the timing breakdowns for writing 1D datasets of 165 ND files (a), 2D datasets of 165 ND files (b), 1D datasets of 6,400 FD files (c), and 2D datasets of 6,400 FD files (d).
Since the size of the  1D datasets is small, we observe poor scalabilities for writing 1D datasets for both ND and FD files.
In contrast, each input ND file has six large 2D datasets of size ranging from 200 MB GB to 6.1 GB.
For FD files, there are five large 2D datasets of size ranging from 53 MB to 7.1 GB.
Concatenating the large 2D datasets for the 165 ND and 6,400 FD files can certainly produce a sufficiently large number of data chunks that they will be distributed among all processes, producing a more balanced workload.
Thus, the timings of writing 2D datasets show more linear scalability as the numbers of processes increase.

We note that the write patterns are different for the two read strategies described in Section \ref{sec:read}.
When using dataset-based partitioning, all the processes collectively read each dataset and write it to the output file one after another.
Thus, the aggregate access file region of each collective write occupies a contiguous file region.
When using file-based partitioning, if a dataset is written in multiple rounds due to the I/O buffer size limit, the aggregate access region of each collective write consists of noncontiguous regions.
The reason is that the entire dataset of an input file must be appended one after another from a different file, and a process's write request in each round covers only partial dataset.
When a larger I/O buffer size is used, fewer noncontiguous collective writes will be required.
We will next discuss the impact of the I/O buffer size on the write performance.

\subsection {Memory Footprint} \label{sec:buffer}

In our study presented so far, we used an I/O buffer large enough to store all the data for the largest dataset in the assigned input files.
If the application's memory space is not large enough, concatenation will have to complete in multiple rounds of read and write, each handling a partial amount of the concatenated dataset.
Thus, the larger the I/O buffer size per process, the fewer the I/O rounds to concatenate a dataset.

In our dataset concatenation case, the I/O buffer size represents the memory footprint of each process, which also determines the number of processes that can run on each compute node without encountering the out-of-memory error.
On Cori, each Haswell node has a memory of size 128 GB.
For example, if each process allocates an I/O buffer of 8 GB, then only up to 16 processes can run on each node.
Because the memory space is used by other programs as well, the actual allowed number of processes per node is lower than 16.
Additionally, if the in-memory I/O feature is enabled for the HDF5 files, the  memory footprint becomes even larger.

For ND files, our experiments show that the maximum I/O buffer size per process to allow the concatenation for a dataset to complete in a single run is 8~GB.
Given the 128 GB memory size on a Haswell node, up to 4 MPI processes per node can be allocated.
If running more processes per node is desired, the I/O buffer size must be proportionally reduced.

Table \ref{table:nproc} presents the maximum numbers of I/O rounds among all datasets for different I/O buffer sizes.
For easy understanding of the impact of the I/O buffer size on the performance, we set the numbers of files and processes to the power of 2.
While keeping  the total number of processes fixed to 128, we vary the number of processes per node between 4 and 32.
As expected, when increasing the number of processes running on each node, the maximum number of I/O rounds increases.
For example, the largest dataset in a single ND file is about 7.2 GB.
If each process allocates an I/O buffer of size 1 GB, then all the processes are required to make 8 collective writes when concatenating that dataset.
For large datasets, using smaller I/O buffer size reduces the memory footprint, but it can increase the number of I/O rounds and degrade the performance of concatenation.

\begin{table}
\scriptsize
\centering
\caption{
    Maximum number of rounds of reads and writes among all the datasets for different I/O buffer sizes.
    The total number of MPI processes and input ND files is fixed at 128, while the number of processes per node varies between 4 and 32.
    The I/O buffer size allocated in each process is 8 GB when running 4 processes per node, and it proportionally decreases as the number of processes run on each node increases.
}
\begin{tabular}{|r|r|r|r|r|r|r|r|} \hline
    \multicolumn{2}{|r|}{Number of processes} & \multirow{2}{*}{4} & \multirow{2}{*}{8} & \multirow{2}{*}{16} & \multirow{2}{*}{32} & \multirow{2}{*}{64} & \multirow{2}{*}{128} \\ \cline{1-2}
    procs per node & I/O buffer & & & & & & \\ \hline \hline
    4 & 8 GB & 24 & 12 & 6 & 4 & 2 & 1 \\ \hline
    8 & 4 GB & - & 24 & 12 & 7 & 4 & 2 \\ \hline
    16 & 2 GB & - & - & 24 & 13 & 7 & 4 \\ \hline
    32 & 1 GB & - & - & - & 25 & 13 & 7 \\ \hline
\end{tabular}
\label{table:nproc}
\end{table}

\textbf{Raw data chunk caching } --
HDF5 supports caching of raw data chunks.
The size of this cache can be adjusted by calling the {\tt H5Pset\_cache} API.
In general, caching can improve performance for data that is repeatedly accessed.
For nonrepeated access patterns, caching can also help performance by enabling I/O aggregation to reduce the number of I/O requests to the file system.

In our case study, the concatenation operation accesses all the datasets no more than once.
The only possibility for caching to take effect is if HDF5 can aggregate the raw data chunks into a single collective MPI file write call, with each rank using an MPI derived data type to describe the noncontiguous memory layouts of raw data chunks.
However, the current implementation of HDF5 does not appear to aggregate raw data chunks across more than one dataset, and
we did not observe a significant performance change when increasing the raw data cache size.

%% file: 7_eval.tex
\section {End-to-End Performance Evaluation} \label{sec:eval}

In this section we study the performance scalability using the best HDF5 feature settings found in the preceding sections.
The end-to-end time used in calculating the speedups is measured from the beginning of opening the input files till the end of closing the output file.
In this experiment the input files are evenly distributed to all the processes and opened in POSIX I/O mode.
The metadata is collected by using in-memory I/O.
Chunk size is set to 1 MB for 1D datasets and keeps the same first dimension chunk size of 128 for 2D datasets as the input files.
All the datasets are chunked and compressed with ZLIB default level 6.
Four MPI processes per compute node are allocated to run the evaluation, while increasing the number of nodes.

\begin{figure}
\centering
\includegraphics[width=\columnwidth]{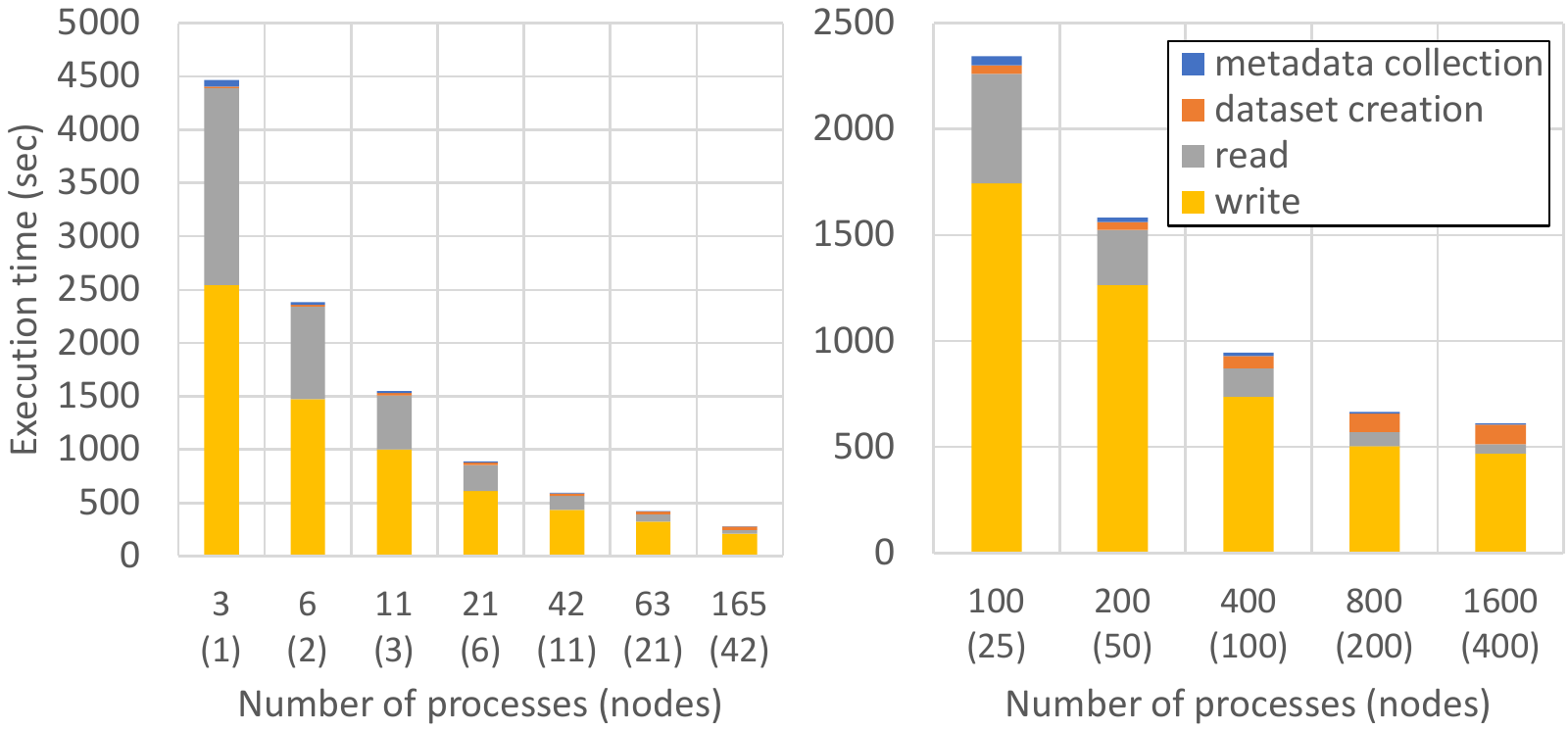}
\caption{
    End-to-end performance of dataset concatenation across 165 ND files (left) and 6,400 FD files (right).
    For ND files, we use up to 165 processes on 42 nodes.
    For FD files, we use up to 1,600 processes on 400 nodes.
}
\label{scale}
\end{figure}

Figure \ref{scale} presents the strong-scaling performance results of the end-to-end dataset concatenation for 165 NOvA ND files (left) and 6,400 NOvA FD files (right).
Note that we ran the smallest case of 100 MPI processes for FD files because any number less than 100 will result in the out-of-memory error.
For 165 ND files, it takes 4467.98 seconds when running 3 processes on a single node.
When it scales up to 165 processes on 42 nodes, the execution time is reduced to 279.4 seconds.
For 6,400 FD files, the dataset concatenation takes 2343.13 seconds when running 100 processes on 25 nodes.
With our fine-tuned HDF5 feature settings, it scales up to 1,600 processes on 400 nodes taking 611.79 seconds to concatenate all the 6,400 HDF5 files.

Table \ref{table:summary1} summarizes the experimental results and the data statistics of the output file.
We see that the output file size is much smaller than the sum of all the input file sizes shown in Table \ref{table:statistics}.
In addition, the metadata in the output file is significantly smaller than the sum of metadata of all the input files.
These differences show the effectiveness of tuned chunk dimension sizes.

\begin{table}
\scriptsize
\centering
\caption{
    Performance summary and data statistics of the largest end-to-end runs for concatenating the ND and FD files.
    With the tuned chunk size, the compression ratio is significantly improved over the sum of the input files, as shown in Table \ref{table:statistics}.
    The overall metadata size is also reduced.
}
\begin{tabular}{|c||c|c|} \hline
    Data & 165 ND files & 6,400 FD files \\ \hline \hline
    Number of processes (nodes) & 165 (42) & 1600 (400) \\ \hline
    Timing & 279.4 sec & 611.8 sec \\ \hline
    Data size before compression & 1 {\bf TB} & 16.9 {\bf TB} \\ \hline
    Output file size & 15.1 GB & 77.9 GB \\ \hline
    Metadata size  & 57.7 MB & 93.8 MB \\ \hline
    Raw data size & 15.0 GB & 77.4 GB \\ \hline
\end{tabular}
\label{table:summary1}
\end{table}

%% file: 8_summary.tex
\section {Summary of HDF5 Feature Tuning} \label{sec:summary}

We have discussed various HDF5 features and their impacts on the performance of the parallel datasets concatenation.
Below is a list of such features used in our study that yield noticeable improvement.
\begin{itemize}
    \item {POSIX I/O mode for reading input files that are disjointly partitioned among processes}
    \item {In-memory I/O for metadata collection}
    \item {Collective dataset creation}
    \item {Chunk size set to 1 MB-based for 1D datasets and 128 row-based ($2$ MB $\sim 44$ MB) for 2D datasets}
    \item {Compression level between $4$ and $6$}
    \item {Increase in the metadata cache size to 128 MB}
    \item {Increase in the metadata block size to at least 4 MB}
    \item {Independent POSIX I/O for reads and collective MPI I/O for writes}
\end{itemize}

Balancing the timings and memory footprints is a delicate task.
While using in-memory I/O can significantly improve the metadata collection speed, it can consume a large amount of memory space.
If the number of data objects in a file is not very large and the raw data takes most of the file space, then in-memory I/O may not be feasible.
An ideal solution in HDF5 would be to prefetch the metadata blocks into an internal buffer, so {\tt H5Ovisit} can look up the metadata in the buffer without reading from the file for each data object.

Chunk size is another delicate parameter that can significantly impact the parallel I/O performance and compression ratios.
Chunk size should be sufficiently small to ensure enough chunks to be shared by the processes for better I/O and compression/decompression parallelism.
However, chunk size should not be too small, because that can yield a large number of chunks causing expensive metadata management.
For NOvA files, the chunk size of 1 MB for 1D datasets provided a good trade-off between the workload balance and the metadata operation cost.
Chunk size also should be large enough to achieve a good compression ratio.
In our case study, a chunk size between 4 MB and 64 MB provided a good compression ratio as well as a reasonable metadata operation cost for 2D NOvA datasets.

HDF5 implements several caching mechanisms for both metadata and raw data; and adjusting the metadata cache size, metadata block size, and raw data cache size can effectively improve the performance.
In our case, when using a metadata cache size of 128 MB with the automatic cache size adjustment turned off, we observed that thewrite performance increased by about $25\%$.
However, we observed no noticeable difference when the size of the raw data chunk cache for raw data I/O was increased.
HDF5 does not yet appear to take advantage of opportunities to aggregate chunks of more than one dataset when writing cached dataset chunks.

In addition to these feature settings in HDF5, we found that the application I/O buffer size affects the parallel I/O performance.
If each process allocated a buffer large enough to store a full dataset read from the locally assigned input files, then the dataset can be concatenated in a single round of read and write.
If the buffer size is limited, however, the concatenation of a dataset must be carried out in multiple rounds of I/O, reading and writing partial datasets one at a time.
Depending on the system hardware's available memory space, using large I/O buffers may limit the number of processes running on a single compute node in order to avoid out-of-memory errors.

Moreover, we found that the following two settings have little performance impact in our dataset concatenation case study: metadata collective or independent I/O mode and raw data chunk cache size.
Because  of the expensive compression and decompression costs, the metadata I/O does not take up a large portion of the overall execution time.
Using collective metadata I/O mode shows a minor improvement for flushing the cached metadata at file close.
However, we anticipate that HDF5 will soon fix the implementation to make use of MPI derived data types to aggregate multiple write requests.
As for the raw data cache size, because the concatenation workflow in this case study does not access the same data more than once, increasing raw data chunk cache size showed no impact to the performance.

%% file: 9_related.tex
\section{Related Work}
PnetCDF \cite{li2003parallel} is a high-level parallel I/O library that is popularly used in scientific communities.
Currently, PnetCDF does not support data compression.
The NOvA data used in this case study is large ($1 \sim 17$ TB) and highly compressible, requiring data compression.
Thus, PnetCDF is considered to be inappropriate for handling this large-scale HEP data.

Adaptive I/O System (ADIOS) \cite{jin2008adaptive} is another high-level parallel I/O library used in scientific communities.
ADIOS supports only BP file format that does not store the data in in their canonical order.
Specifically, in a parallel write operation, ADIOS simply appends the data from one process to the data from another process.
In addition, there are not many third party software that support I/O to BP format.
Given its rich third-party software ecosystem, HDF5 has become the primary choice for data storage and parallel I/O explorations within the HEP community.
The NOvA researchers chose HDF5 because of this exact reason.

To the best of our knowledge, there is only a limited number of published papers that discuss the parallel I/O performance of HDF5 when the data compression feature is enabled.
While they show a performance comparison across five different compression settings, they only consider the parallel write performance for a relatively small amount of data (the compressed data size is $2 \sim 3$ GB).
Pokhrel et al. also shows a comparison of parallel I/O performance across different compression algorithms \cite{pokhrel2018parallel} using up to 38 processes only.
Most of the previous works that studied the parallel HDF5 performance using I/O benchmarks, such as IOR \cite{IOR}, do not study the impact of data compression on the parallel I/O performance \cite{shan2008characterizing,xu2016lioprof,wauteleta2011parallel,shan2007using,saini2007parallel}.
Kunkel et al. analyzed the impact of different compression algorithms on the parallel I/O performance using climate data \cite{kunkel2020aimes}.
We present a comprehensive empirical study on the HDF5 parallel I/O performance, especially focused on how to exploit a variety of HDF5 features for the parallel I/O with data compression in this specific HEP data concatenation case.

%% file: 10_conclusion.tex
\section{Conclusion} \label{sec:conclusion}

HEP experimental data is typically stored in a large number of files based on their chronological creation order.
Concatenating this data is the first step for many HEP analysis programs.
In this paper we have investigated a variety of HDF5 features and analyzed their impact on dataset concatenation performance using experimental NOvA data files.
We studied how to tune various HDF5 features to achieve scalable parallel I/O performance over a large number of chunked, compressed datasets.

Currently, HDF5 has some limitations on collective I/O for compressed datasets.
For example, HDF5 allows only a single dataset to be collectively accessed at a time.
We believe that the parallel I/O performance could be improved by relaxing such a limitation, especially when concatenating a large number of small datasets.
The lessons learned from our case study can provide guidance in selecting the appropriate HDF5 parameters for scientific applications that exhibit similar I/O characteristics, such as a large number of datasets per file and variable sizes of the datasets in each file.